\documentclass[prl,twocolumn,showpacs]{revtex4}
\usepackage{bm}
\usepackage{graphicx}
\usepackage{amssymb}
\usepackage{amsmath}
\usepackage{ulem}
\usepackage{color}
\begin{document}
%%%%%%%%%%%%%%%%%%%%%%%%%%%%%%%%%%%%%%%%%%%%%%%%%%%%%%%%%%%%%%%%%%%%%%%%%%%%%%%%%%%%
\title{{High intensity study of THz detectors based on field effect transistors}}
%%%%%%%%%%%%%
\author
{D. B. But$^{1,2}$, C. Drexler$^3$, M. V. Sakhno$^2$, N. Dyakonova$^1$ , O. Drachenko$^4$, F. F. Sizov$^2$, A. Gutin$^5$, S. D. Ganichev$^3$, W. Knap$^1$}

\affiliation{$^1$ UMR 5221 CNRS, Universite´ Montpellier 2, Montpellier 34095, France}

\affiliation{$^2$ V.E. Lashkaryov Inst Semicond Phys, Kiev, 03028, Ukraine}

\affiliation{$^3$ Terahertz Center, University of Regensburg, Regensburg, 93040, Germany}

\affiliation{$^4$ Helmholtz Zentrum Dresden Rossendorf, Inst Ion Beam Phys $\&$ Mat Res, Dresden, 01314, Germany}

\affiliation{$^5$ Rensselaer Polytechnic Institute, Troy, New York, 12180, USA}
%%%%%%%%%%%%
\begin{abstract}
Terahertz power dependence of the photoresponse of field effect transistors, operating at frequencies from 0.1 to 3~THz for incident radiation power density up to 100 kW/cm$^2$ was studied for Si metal-oxide-semiconductor field-effect transistors and InGaAs high electron mobility transistors. The photoresponse increased linearly with increasing radiation power up to kW/cm$^2$ range. The saturation of the photoresponse was observed for all investigated field effect transistors for intensities above several kW/cm$^2$. The observed signal saturation is explained by drain photocurrent saturation similar to saturation in direct currents output characteristics. The theoretical model of terahertz field effect transistor photoresponse at high intensity was developed. The model explains quantitatively experimental data both in linear and nonlinear (saturation) range. Our results show that dynamic range of field effect transistors is very high and can extend over more than six orderd of magnitudes  of power densities (from 0.5~mW/cm$^2$ to 5~kW/cm$^2$).
\end{abstract}
%%%%%%%%%%%%
\date{\today}
\maketitle
%%%%%%%%%%%%
\section{Introduction}
%%%%%%%%%%%%
Electromagnetic waves in the terahertz (THz) frequency range are gaining in importance because of many applications in domains of security, biology, imaging, material control, characterization, etc.  The development of many of these applications is hindered by the lack of sensitive, room temperature robust detectors, especially with high dynamic range.  The THz detection phenomenon in field-effect transistors (FET) was explained by the Dyakonov-Shur plasma wave theory~\cite{bib41}.  When THz radiation is coupled to the FET – between gate and source the THz \textit{ac} voltage modulates simultaneously carrier density and the carrier drift velocity.  As a result THz \textit{ac} signal is rectified and leads to a \textit{dc} photoresponse proportional to the received power.  For high carrier mobility devices (III-V devices at cryogenic temperatures) the THz field can induce plasma waves that propagate in the channel and resonant plasma modes can be excited leading to very efficient  narrowband and voltage tunable detection~\cite{bib02, bib42}.  At room temperature plasma waves are usually over damped and THz radiation leads only to a density perturbation that decays exponentially with the distance from the source (drain) with a characteristic length $L_{eff}$ that is typically of the order of a few tens of nanometers~\cite{bib18}.  A more detailed description of the physical mechanism of THz detection by FETs can be found in Ref.~\cite{bib43}. In the case of room temperature broadband detection (over damped plasma) the detection process can be alternatively explained by the model of distributed resistive self-mixing~\cite{bib30, bib33}.  Although not treating all plasma related physics rigorously, the resistive mixing model allows a rational detector design~\cite{bib33, bib44}.  THz FET detectors show high responsivity (up to a few kV/W), low noise equivalent power (down to 10 pW$/\sqrt{\text{Hz}}$)~\cite{bib03} and fast response time (below 1~ns)~\cite{bib04} and (below 30~ps)~\cite{bib05}.  Further interesting property of the FET detector was recently discovered: the photoresponse exhibits sensitivity to the radiation helicity~\cite{bib04, bib06} making FET promising for the all-electric detection of the radiation Stokes parameters.  The combination of fast response and high sensitivity makes FETs also promising detectors for monitoring many THz sources.  However, there are very few studies of these detectors dynamic range -- especially at high intensities.  The first high intensity experiments were reported by Preu, et al.~\cite{bib07}.  It was shown that for GaAs high electron mobility transistors (HEMTs) the linearity of the photoresponse holds up to 8 $\pm$ 4~kW/cm$^2$ (11~W) of beam intensity at 240~GHz and the linear region is followed by the sub-linear one.  For moderate intensity (up to 510~W/cm$^2$ at 1.63~THz) the broadband detection by FETs versus the radiation power was studied by Gutin, et al.~\cite{bib08}, who observed linear region, followed by square root dependence for higher radiation intensities.

Here we report on the observation of the radiation intensity dependent photoresponse of FETs in a very wide intensity range going up to 500~kW/cm$^2$.  We demonstrate that for a InGaAs HEMTs and silicon metal-oxide-semiconductor field-effect transistors (Si-MOSFETs), the photoresponse has a substantial linear dependence followed by the nonlinear region and saturation.  The model of THz FET photoresponse working in a wide range of intensities  was developed. It allows quantitative data interpretation using parameters determined from direct currents (\textit{dc}) output characteristics.  The calculated THz FET photoresponse describes well experimental data both in linear and nonlinear (saturation) range.  Our experimental results show that dynamic range of FETs based THz detectors is relatively high and can extend over a wide range of intensities from $\sim$~0.5~mW/cm$^2$ to 5~kW/cm$^2$.
%%%%%%%%%%%%
\section{Experiment}
%%%%%%%%%%%%
To study the detectors responsivity in a wide intensity range from 0.5~mW/cm$^2$  to 500~kW/cm$^2$  we used several types of monochromatic continuous waves (\textit{cw}) and pulsed sources of terahertz radiation operating in the frequency range from 0.13~THz up to 3.3~THz.

Low power \textit{cw} radiation ($<$1~W/cm$^2$ ) was obtained applying backward wave oscillators (BWO), the \textit{cw} methanol laser~\cite{bib09, bib10} and the commercially available Schottky diode (Radiometer Physics GmbH).  The operation frequencies and maximum output power levels were as follows: 129-145~GHz and 15~mW (BWO, $\Gamma $4-161), 2.54~THz and 20~mW (methanol laser), and 292~GHz and 4.5~mW (Schottky diode source, Radiometer Physics GmbH).  Radiation from the sources was focused onto the sample; beam spatial distribution was monitored during the experiment.  Radiation attenuation was controlled by a set of calibrated attenuators or directly by changing a source output power.  Detection signal was measured using a lock-in amplifier technique.
%%%%%%%%%%%%FIG.1%%%%%%%%%%%%
\begin{figure}
\includegraphics[height=0.3\linewidth]{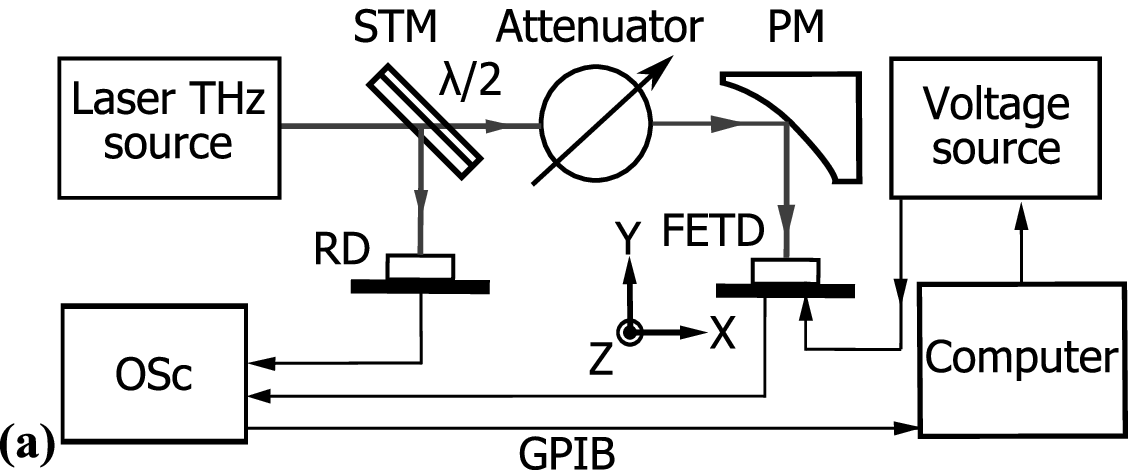}
\includegraphics[height=0.3\linewidth]{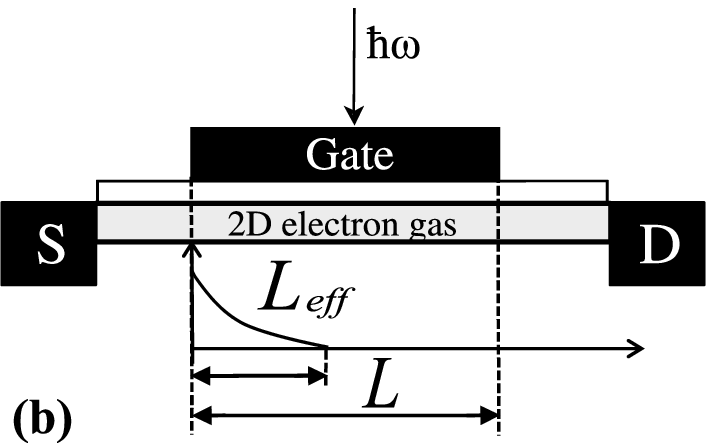}
\caption{(a) Schematic illustration of experimental setup. STM is a semitransparent mirror, PM is a parabolic mirror, OSc is an oscilloscope, RD is a reference detector, FETD is THz FET detector. (b) Schematic view of the HEMT device with contact terminals S (source), D (drain), and G (gate). Transistors were irradiated by linearly polarized radiation at normal incidence in all optical experiments. \textit{L} and $L_{eff}$ are length and effective detection length of transistor channel, respectively.}
\label{Fig:fig01}
\end{figure}
%%%%%%%%%%%%

As sources of high power radiation we used optically pumped molecular terahertz laser of the Regensburg Terahertz Center TerZ~\cite{bib11, bib12, bib10} and free electron laser at Rossendorf Helmholz Institute system FELBE~\cite{bib14}.  Using NH$_3$ as active medium we obtained linearly polarized radiation with frequencies of 3.33, 2.03 and 1.07~THz.  Lower frequencies of 0.78 and 0.61~THz were achieved using D$_2$O, and CH$_3$F, respectively.  The molecular laser generated single pulses with duration of about 100~ns, peak intensity of $\sim$~500~kW/cm$^2$, and a repetition rate of 1~Hz.  The radiation power was controlled by the THz photon drag detector~\cite{bib15} (Fig.~\ref{Fig:fig01}(a)).  By focusing the laser beam with a parabolic mirror we achieved an almost Gaussian profile, as recorded with a pyroelectric camera~\cite{bib16}, and exhibiting full widths at half maximum between 1~mm (at 3.33~THz) and 3~mm (at 0.61~THz).  To vary the radiation intensity we used a set of teflon, black polyethylene and/or pertinax calibrated attenuators~\cite{bib17}.  In this set-up the photoresponses were picked up as a voltage drop over a 50~$\Omega$ load resistor and fed into an amplifier with a bandwidth of 300~MHz and a voltage amplification of 20~dB.  The FEL provided 6.22~ps pulses at 1.55~THz and 8.4~ps pulses at 2.11~THz with a repetition rate of 13~MHz, peak intensity of $P~\sim$~200~kW/cm$^2$.  The radiation attenuation was achieved using a series of wire grid attenuators.  The beam was focused on the sample by a parabolic mirror; the resulting diameter was about 2.5~cm.

The main parameters of THz FETs detectors used in our experiments and some simulations parameters are summarized in Table~\ref{tab:tab01}.  Our experiments were done at room temperature, thus the regime of the detection was non-resonant (broadband).  The radiation induced carrier density oscillations exist only near the source (or drain) side of the channel $L_{eff}$~\cite{bib18} where the radiation is fed (see Fig.~\ref{Fig:fig01}(b)).
\begin{widetext}
%%%%%%%%%%%%Table I%%%%%%%%%%%%
\begin{table}[h]
%\tiny   %%%%%%%%%%%%-TABLE SIZE-%%%%%%%%%%%%
\caption{Sample parameters} %title of the table
%\centering % centering table
\begin{tabular}{c c c c c c c c c c c c c} % creating eight columns
\hline\hline
Name& Type & \textit{L} & \textit{W} & $V_{th}$ & $\eta$ & $\mu_{n}$ & $k_{th}$ & $k_{\mu}$ & $\alpha $ & $\beta$ & $L_{eff}$ & $k_{ant}$\\ [0.5ex]
\hline
 %&  & $\mu$m & $\mu$m & V &  & $\frac{cm^2}{V•s}$ & $\frac{mV}{K}$ &  &  &  & nm &  \\ \hline
  &  & $\mu$m & $\mu$m & V &  & cm$^2$/V$\cdot$s & mV/K &  &  &  & nm & cm $\cdot \sqrt{\text{V}/\text{W}}$ \\ \hline
HEMT & InGaAs & 0.13 & 12-40 & -0.18 & 1.3 & 2900 & 0.21 & 1.6 & 0.03 & 0.9 & 65 & 212 \\ \hline
MOSFET & Si & 2 & 20 & 0.6 & 1.75 & 500 & 2.1 & 1.7 & 0.012 & 0.9 & 31 & 298 \\ %[1ex]
\hline
\hline
\end{tabular}
\label{tab:tab01}
\end{table}
%%%%%%%%%%%%END%%%%%%%%%%%%
\end{widetext}

The HEMT transistors were pseudomorphic HEMTs ones based on InGaAs/GaAs structures.  The gate length was \textit{L}~=~0.13~$\mu$m, the gate width was from 10 up to 40~$\mu$m,  open channel carriers mobility was \textit{$\mu_n$}~=~2900~cm$^2$/(V•s) at \textit{T}~=~300~K. The carrier mobility was \textit{$\mu_n$}~=~1200~cm$^2$/(V•s) near the threshold voltage (closed channel).  The mobility was determined using magneto-resistance technique~\cite{bib19}.  We studied also Si~MOSFETs: \textit{L}~=~2~$\mu$m, \textit{$\mu_n$}~=~500~cm$^2$/(V•s).  These transistors did not comprise any specially designed antenna for the incident radiation coupling.  Similarly to other experiments the bonding wires and metallization of contact pads served as effective antennas~\cite{bib20, bib21}.
%%%%%%%%%%%%
\section{Results}
\begin{figure}
\includegraphics[width=0.8\linewidth]{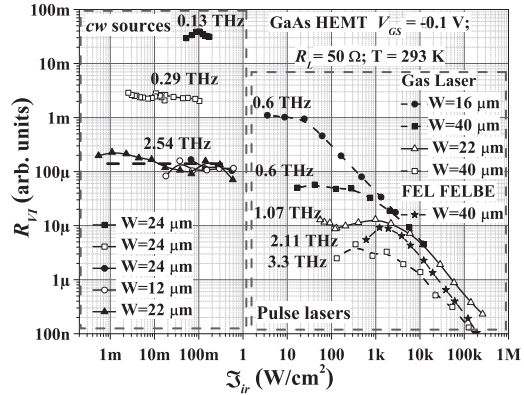}
\caption{The responsivity \textit{R$_{VI}$} of several HEMTs for different intensity and frequency. By using the \textit{cw} sources we observed only the linear regime of \textit{R$_{VI}$}(\textit{$\Im_{ir}$}) independently of their radiation frequency(on the left). Only high intensity pulsed lasers permitted to observe the limit of FET detection (on the right). Saturation intensity increases with frequency.}
\label{Fig:fig02}
\end{figure}

%%%%%%%%%%%%
Figure~\ref{Fig:fig02} shows compilation of the results for different HEMT samples, different types of THz sources operating in a wide intensity range from 0.5~mW/cm$^2$ to 500~kW/cm$^2$ and at different frequencies in the range from 0.13~THz up to 3.3~THz.  The responsivity value $R_{VI}$ is defined as the ratio of THz FET detector signal $\Delta U$ to radiation intensity $\Im_{ir}$ in focal point of lens (or parabolic mirror) and is plotted as a function of power density on the detector.  Figure~\ref{Fig:fig02} is separated into two groups.  The left group in the figure shows responsivity for \textit{cw} sources with low output intensity (or power less than a few~mW).  The right group shows responsivity for high intensity THz pulse sources.  As can be seen in Fig.~\ref{Fig:fig02}, the responsivity values of THz FETs in pulse lasers group are less than responsivity values in \textit{cw} sources.  We suppose difference in responsivity can be attributed to a responsivity limiting by additional capacitances of the readout circuit in pulse experiments.

The obtained result could be fitted by the phenomenological formula for determining of the characteristic saturation intensity $\Im_{ir,sat}$:
\begin{equation}
R_{VI} (\Im_{ir})=R_{VI0} \frac{1}{1+\Im_{ir}/\Im_{ir,sat}} ,
\label{Eq:eq3}
\end{equation}

where, $R_{VI0} = {\Delta U}/\Im_{ir }$ is the constant responsivity in the linear regime.
%%%%%%%%%%%%FIG.2%%%%%%%%%%%%
%%%%%%%%%%%%%%%%%%%%%%%%

The data in Fig.~\ref{Fig:fig02} show that after the constant value (linear range) $R_{VI}$ decreases and characteristic saturation intensities shift to higher intensity values when frequency $\omega$ increases.  The frequency dependence of $\Delta U$ in linear region originates from combined effects of the frequency dependent properties of both the device and the antenna~\cite{bib22, bib23, bib24}.  The $R_{VI}$ for low intensity linear regime can be written as:
\begin{equation}
R_{VI}(\Im_{ir},\omega_0)=R_{VI} (\Im_{ir},\omega_0)(\frac{\omega_0}{\omega})^{\gamma},
\label{Eq:eq1}
\end{equation}

\begin{figure}
\includegraphics[width=0.8\linewidth]{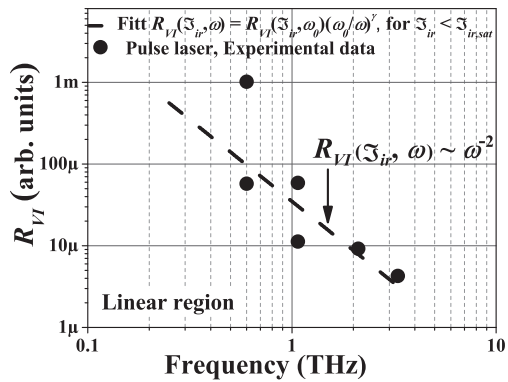}
\caption{Experimental data of THz FET responsivity $R_{VI}(\Im_{ir},~\omega)$ versus frequency in the linear region for pulsed detection regime (open dots) and $1/\omega^2$ slope (dashed line).}
\label{Fig:fig03}
\end{figure}

The influence of THz FETs parameters on photoresponse and their matching with the incident radiation for the low input powers have been analyzed in details in Ref.~\cite{bib24}.  It was shown that \textit{$\gamma$}~=~2 (Eq.~\eqref{Eq:eq1}), if lenses or wide aperture antennas are used in experiments, and \textit{$\gamma$}~=~4 for other cases at \textit{f} $\geq$ 1~THz.  Our experimental data obtained in linear regime (Fig.~\ref{Fig:fig03}) agree with the behavior of responsivity which was proposed in Ref.~\cite{bib24} (\textit{$\gamma$}~=~2), and are similar to the data given in Ref.~\cite{bib23}.
%%%%%%%%%%%%FIG3%%%%%%%%%%%%

Figure~\ref{Fig:fig03} shows frequency dependency of \textit{R$_{VI}$} which was obtained in the linear regime of signals in pulse laser group at 10~W/cm$^2$ (after Fig.~\ref{Fig:fig02}).  This result shows that $\gamma$~=~2 and allows us to estimate frequency behavior.

\begin{figure}
\includegraphics[width=0.8\linewidth]{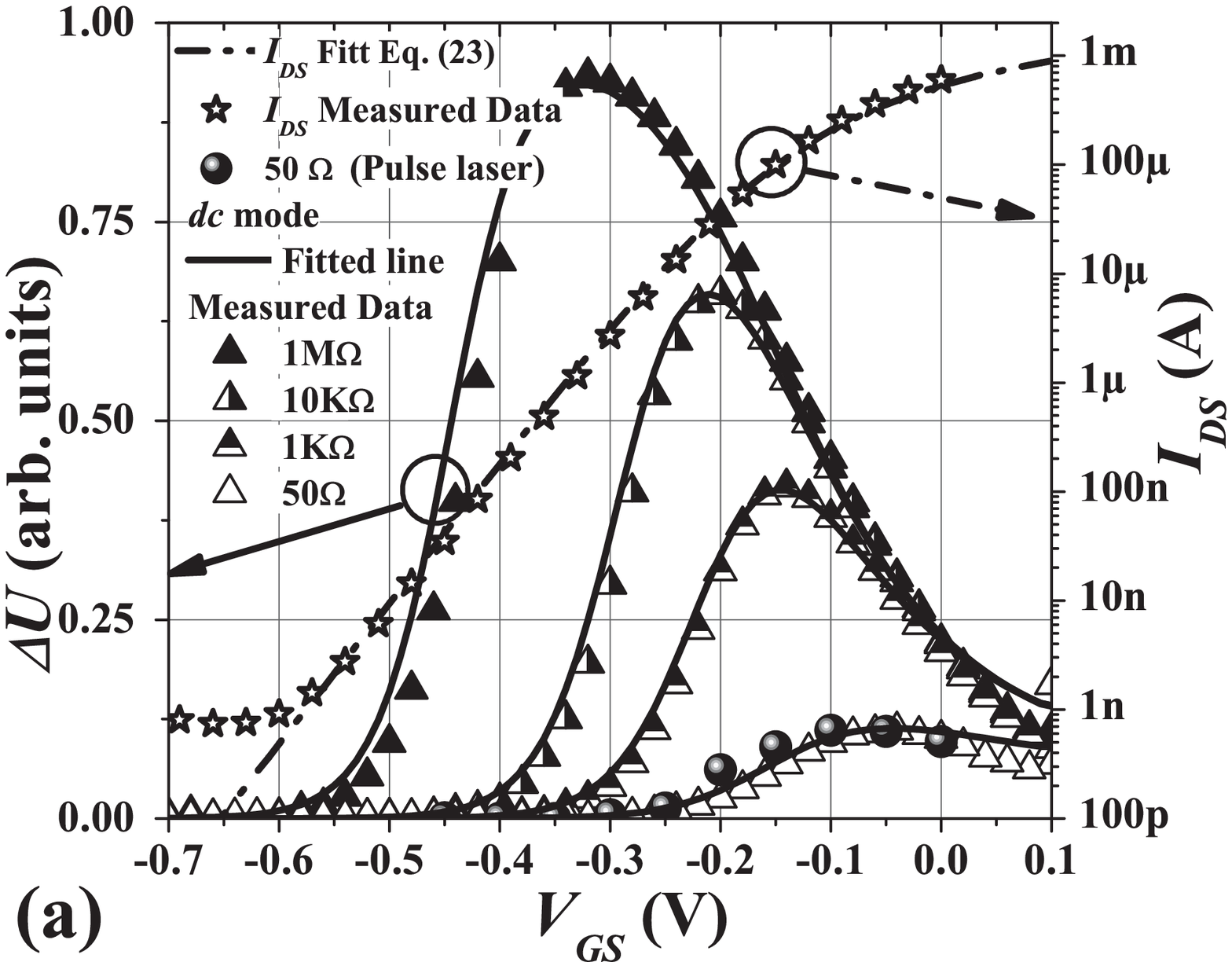}
\includegraphics[width=0.8\linewidth]{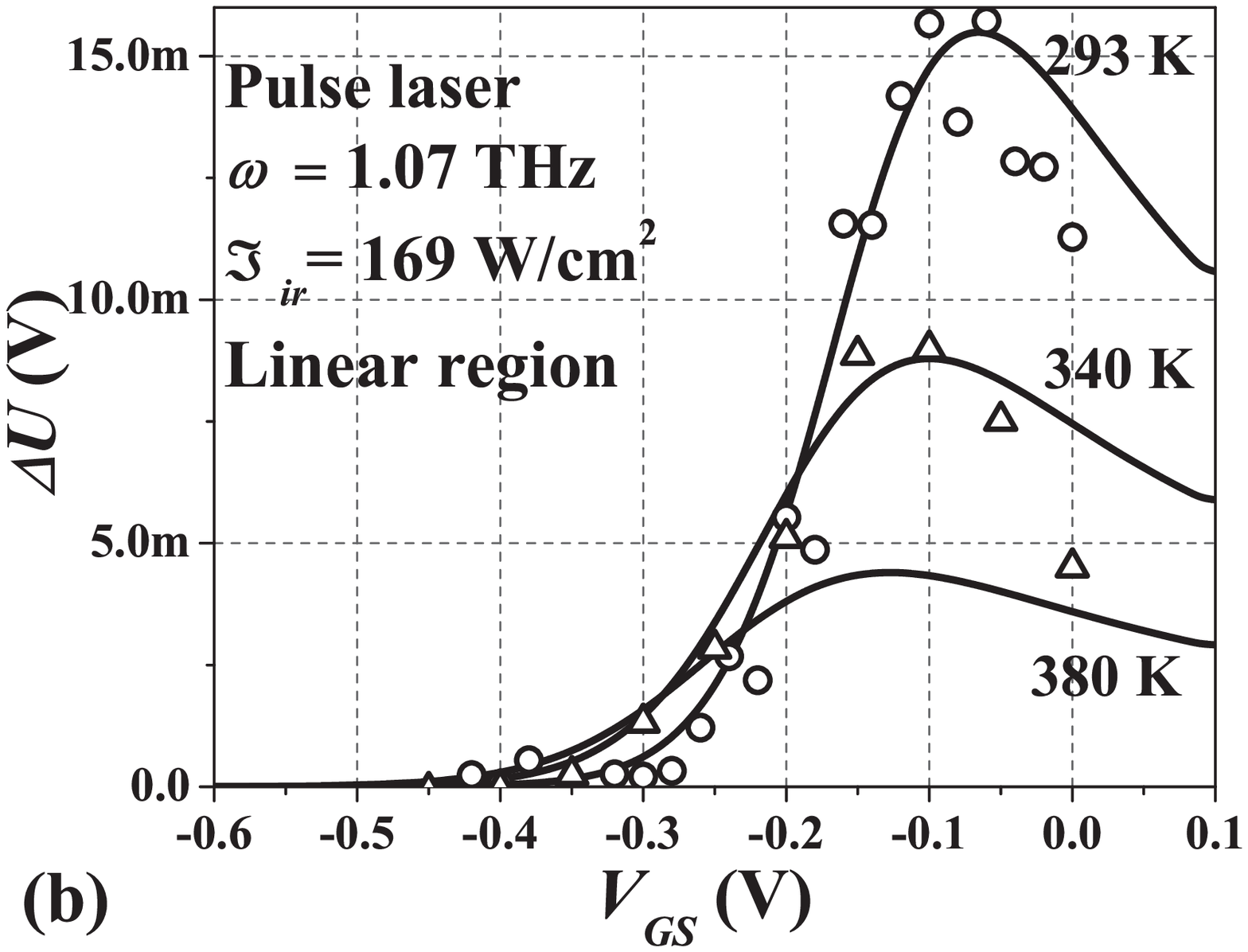}
\caption{(a)~Left ordinate: FET photoresponse to \textit{cw} THz radiation (triangles) measured as a function of gate bias at $R_L$ 1~M$\Omega$, 10~k$\Omega$, 1~k$\Omega$ and 50~$\Omega$. Photoresponse to pulsed radiation measured across 50~$\Omega$ is shown by full circles. Here the photoresponse was obtained in low radiation intensity regime. Right ordinate: dark drain current as a function of gate bias: experimental results (stars) and fit (dash-dot line) according to Eq.~\eqref{Eq:eq13} using parameters of HEMT presented in Table~\ref{tab:tab01}. (b)~Comparison of photoresponses at different temperatures for HEMT: circles and rectangles are experimental data, lines are calculated using Eq.~\eqref{Eq:eq20} and taking into account the temperature dependence Eq.~\eqref{Eq:eq22} and Eq.~\eqref{Eq:eq23} (for parameters $k_{th}$, $k_{\mu}$, see Table~\ref{tab:tab01})}
\label{Fig:fig04}
\end{figure}

Figure~\ref{Fig:fig04}(a) presents photoresponses $\Delta U$ (left ordinate) to \textit{cw} THz radiation (triangles) measured as a function of gate bias at different loading resistances in low radiation intensity regime and constant radiation frequency.  The response time of the transistor is determined by the time resolution of our setup, but it is 2~ns or less.  Thus, we measured the signals at 50~$\Omega$ loading resistance for using the pulsed radiation sources (see Fig.~\ref{Fig:fig01}(a)).  This permitted to obtain identical $\Delta U({V_{GS}})$ dependences with pulsed and \textit{cw} sources (compare empty triangles and circles in Fig.~\ref{Fig:fig04}(a)).  In this case, the FET was in the strong inversion regime (above-threshold) and the photoresponse can be described by the model (see Eq.~\eqref{Eq:eq8} below) at constant $\omega$ and radiation intensity.
%%%%%%%%%%%%FIG4%%%%%%%%%%%%

Figure~\ref{Fig:fig04}(b) shows photoresponse as a function of gate voltage and the decreasing of $\Delta U$ due to temperature rise for HEMT (\textit{W}~=~24~$\mu$m) in linear region of photoresponse. Solid lines in Fig.~\ref{Fig:fig04}(b) presents $\Delta U$ calculated using model (Eq.~\eqref{Eq:eq8}) taking into account heating of samples (Eq.~\eqref{Eq:eq22} and Eq.~\eqref{Eq:eq23}).  Maximum of photoresponse swings to more negative gate voltage ($V_{th}$ =~-0.19~V at \textit{T} =~293~K and $V_{th}$~=~-0.225~V at 340~K) and photoresponse shape become flatter with maximum value reducing ($\Delta U$~=~15.5~mV at \textit{T}~=~293~K and $\Delta U$~=~8.8~mV at 340~K) with temperature rise.

\begin{figure}
\includegraphics[width=0.8\linewidth]{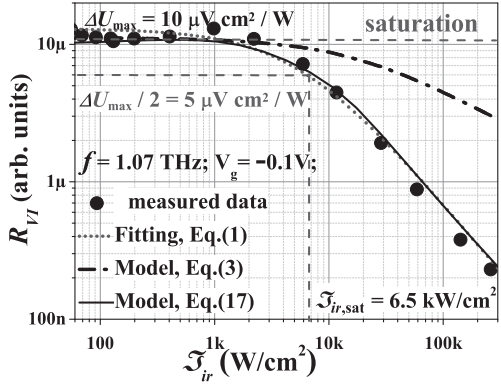}
\caption{Responsivity $R_{VI}$  of HEMT sample ($W$~=~22~$\mu$m) as a function of radiation intensity. Points are the experimental data, dashed line is the theoretical prediction~\cite{bib08} Eq.~\eqref{Eq:eq2}, points line is the fit of Eq.~\eqref{Eq:eq3}; solid line is $R_{VI}$ using the model Eq.~\eqref{Eq:eq21} which takes into account the nonlinear behavior of the photocurrent in the transistor channel. As antenna parameters were unknown we multiplied the model results by a constant to match with linear region of experiment.}
\label{Fig:fig05}
\end{figure}

At higher intensities the photoresponse is no longer proportional to the incoming power (see Fig.~\ref{Fig:fig05}).  Sub-linear dependence was demonstrated in experiments~\cite{bib07} on GaAs HEMTs. The linearity of the photoresponse was observed up to 8~$\pm$~4~kW/cm$^2$ (11~W) of FEL (UCSB, Santa Barbara) beam power at 240~GHz, (the authors suggested that Si~lenses focused from 10\% up to 50\% of incident power on the devices) and the nonlinear region is followed by saturation.  The theoretical model in Ref. Gutin, et al.~\cite{bib08} based on model~\cite{bib25, bib26} considerations provides analytical expressions for two cases of gate voltage: above and below threshold voltage.  In our case, when FETs operates above threshold regimes ($V_0=V_{GS}-V_{th}>$0) the theory~\cite{bib08} gives the following expression for the photoresponse:
\begin{equation}
\Delta U = \frac{U_\text{a}^2}{2 \sqrt{V_0^2+U_{\text{a}}^2/2}+V_0^2} ,
\label{Eq:eq2}
\end{equation}

It was found that at high input radiation signals the response $\Delta U$ becomes non-linear function of the intensity, namely a square root function.  In Ref.~\cite{bib08}, experimental result at 1.63~THz with power level varying from 9.5~W/cm$^2$ up to 510~W/cm$^2$ (after paper 3~mW to 160~mW at beam diameter was 0.2~mm) indeed showed slightly sub-linear behavior in agreement with the theory~\cite{bib08}.  Here, at frequencies 1.07~THz we observed the linear dependence up to intensity of $\sim$~6.5~kW/cm$^2$.  However, our results (full points in Fig.~\ref{Fig:fig05}) show a saturation behavior and can not be described by square root dependence proposed by Eq.~\ref{Eq:eq2} (the dash dot line in Fig.~\ref{Fig:fig05}).  This model (Eq.~\ref{Eq:eq2}) can describe only a small part of signal \textit{vs} intensity after linear regime.
%%%%%%%%%%%%FIG5%%%%%%%%%%%%

\begin{figure}
\includegraphics[width=0.8\linewidth]{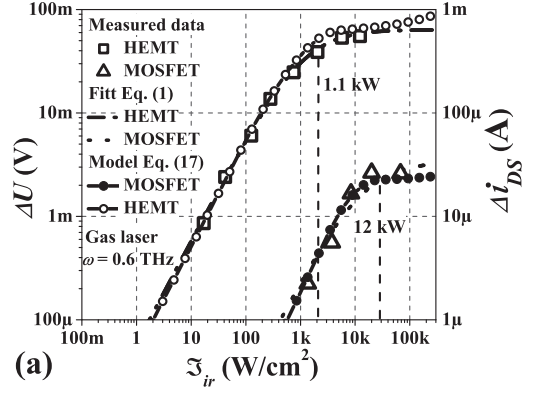}
\includegraphics[width=0.8\linewidth]{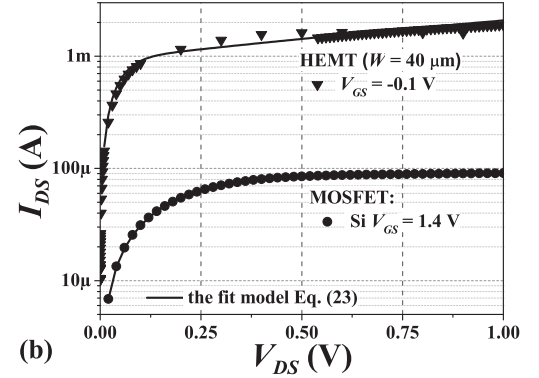}
\caption{(a) Photoresponse of FETs based on InGaAs (HEMT $W$~=~40~$\mu$m) and Si (MOSFET in Table~\ref{tab:tab01}) as a function of radiation intensity. The radiation frequency is 0.6~THz, gate voltages for HEMT is $V_{GS}$ = -0.1 V, $V_{GS}$ = 1.4 V for MOSFET. Squares and triangles are experimental data. Dashed lines are fit according to phenomenological Eq.~\eqref{Eq:eq3}. Dot-lines are fit according to model Eq.~\eqref{Eq:eq21} using parameters of Table~\ref{tab:tab01}. (b) Dots are experimental data of FETs output characteristics at work point of gate voltage $V_{GS}$, Lines are fit according to Eq.~\eqref{Eq:eq13} using parameters of Table~\ref{tab:tab01}.}
\label{Fig:fig06}
\end{figure}

Photoresponse $\Delta U$ of InGaAs~HEMT and Si~MOSFET to pulsed radiation is shown in Fig.~\ref{Fig:fig06}(a). Although the photoresponse between transistors differs significantly, the dependence of photoresponse on the radiation intensity is qualitatively the same: at first it is linear then it saturates.
%%%%%%%%%%%%FIG6%%%%%%%%%%%%

The saturation values of photoresponse $\Delta U_{sat}$ in Fig.~\ref{Fig:fig06}(a) are 70~mV for HEMT and 3~mV for MOSFET.  We convert $\Delta U_{sat}$ into the current $\Delta i_{DS,sat}$ in measured circuit as ($\Delta i_{DS,sat}$~=~$\Delta U_{sat}/(R_L+R_{ch}$), $R_L$, $R_{ch}$ are load and transistor channel resistance, respectively) and obtain 0.7~mA and 30~$\mu$A (see the right ordinate axis in Fig.~\ref{Fig:fig06}(a).  It is important to note that these values are close to saturation currents in the dark at the same gate bias as it is seen from Fig.~\ref{Fig:fig06}(b).  $I_{DS,sat} \approx$~1~mA for HEMTs (see Fig.~\ref{Fig:fig04}(a)) and $I_{DS,sat}$~=~90~$\mu$A for MOSFET. This important observation will be discussed below.
%%%%%%%%%%%%
\section{Broadband detection model and discussion}
%%%%%%%%%%%%
The transistor converts high frequency voltage $V_{tr}$ into \textit{dc} voltage $\Delta U$ which is measured.  Antenna of THz FETs detectors transforms incoming irradiation into voltage $V_{ant}$.  Voltage $V_{tr}$ on FET THz detector is given by:
\begin{equation}
V_{tr} = \eta_{ant} V_{ant},
\label{Eq:eq4}
\end{equation}

where, $\eta_{ant}$ is the coefficient representing losses in the parasitic elements and impedance mismatch between antenna and FET.  The input impedance of the internal transistor part is defined by the input impedance of a transmission line.  The input impedance is replaced by the factor $\eta_{ant}$ in Eq.~\eqref{Eq:eq4}, since the determination of input impedance is a difficult experimental task.

The maximum power $P_{ant}$, which is supplied to FET channel from antenna at plane illumination, is given by~\cite{bib27}:
\begin{equation}
P_{ant} = G\frac{\lambda^2}{4\pi} \Im_{ir} ,
\label{Eq:eq5}
\end{equation}

here, \textit{G} is antenna gain coefficient, $\lambda$ is a wavelength in vacuum, $\Im_{ir}$ is a incoming radiation intensity.  The $V_{ant}$ can be written using the last Eq.~\eqref{Eq:eq5}:
\begin{equation}
V_{ant}^2 = G\frac{2}{\pi} \lambda^2 \Im_{ir} \text{Re}Z_{ant},
\label{Eq:eq6}
\end{equation}

Taking into account Eq.~\eqref{Eq:eq6} the Eq.~\eqref{Eq:eq4} can be rewritten as:
\begin{equation}
V_{tr} = k_{ant}\sqrt{\Im_{ir}} ,
\label{Eq:eq7}
\end{equation}

where, $k_{ant}$ is a parameter which depends on antenna parameters and impedance matching.

The effective rectification length $L_{eff}$ is estimated from the condition at which THz voltage signal is decreased by \textit{e} times along the channel (see Fig.~\ref{Fig:fig01}(b)) and can be approximately calculated as~\cite{bib26, bib31}:
\begin{equation}
L_{eff}=\sqrt{\frac{\mu_{n}n}{\omega (dn/dU)|_{U=V_{GS}}}} \approx \sqrt{\frac{\mu_{n} (V_{GS}-V_{th}))}{\omega}},
\label{Eq:eq14}
\end{equation}

where, \textit{n} is the electron concentration in channel.

Estimates of samples $L_{eff}$ are shown in Table~\ref{tab:tab01} for 0.6~THz.
%%%%%%%%%%%%FIG7%%%%%%%%%%%%
\begin{figure}
\includegraphics[width=0.8\linewidth]{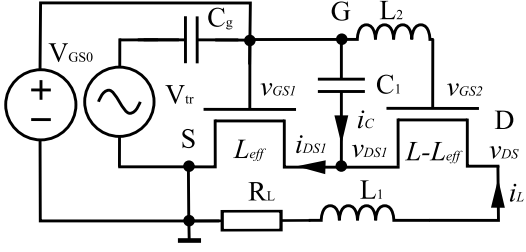}
\caption{The scheme of THz detection by FET with load parasitic elements. S, D, G are the source, drain and gate terminals of FET THz detector, respectively; $R_L$ is the resistance of measurement circuit, \textit{L} and $L_{eff}$ are length and effective detection length of transistor channel, respectively, $i_L$ is photocurrent, $V_{tr}$ is the amplitude of the external \textit{ac} voltage, $V_{GS0}$ is the external \textit{dc} voltage on gate.}
\label{Fig:fig07}
\end{figure}

As was shown by~\cite{bib32,bib33} the transistor rectifying THz radiation can be represented by distributed \textit{RLC} elements.  Our model is based on assumption that FET can be divided into two parts: the first one with length $L_{eff}$ where the rectification takes place (see Fig.~\ref{Fig:fig01}(b)) and the second one with length ($L-L_{eff}$) connected in series (see Fig.~\ref{Fig:fig07}).  THz voltage and current are present only in the first part.  The second part is considered like a load part.  To model such behavor we add elements $L_1$, $L_2$, $C_1$ to equivalent circuit.  Inductance $L_1$ models that THz current does not flow through the load.  Inductance $L_2$ models the absence of THz voltage on the gate of the load part with length ($L-L_{eff}$).  Capacitance $C_1$ is the gate-to-channel capacitance of the load part, the $i_c$ represents the \textit{ac} THz current through the gate.  $C_g$ models two sources ($V_{GS0}$,~$V_{tr}$) decoupling.

The photocurrent flows in the circuit formed by the THz FET detector and an external load $R_L$.  The $v_{DS}$ is the voltage between source and drain.  The $v_{DS1}$ the voltage drop on a characteristic length $L_{eff}$ due to the rectification of THz voltage; and $v_{DS2}$ is the voltage drop on the length ($L-L_{eff}$).  Current through part $L_{eff}$ is given by Eq.~\eqref{Eq:eq13} with corresponding values
\begin{equation}
i_{DS1} = I_{DS}(v_{DS1}, v_{GS1}, L_{eff}) ,
\label{Eq:eq15}
\end{equation}
where,
\begin{equation}
v_{DS1} = -v_{DS2}-R_L i_L + V_{tr}\cos{(\omega t)} ,
\label{Eq:eq16}
\end{equation}
\begin{equation}
v_{GS1} = V_{GS0}+V_{tr}\cos{(\omega t)} ,
\label{Eq:eq17}
\end{equation}
Current through part ($L-L_{eff}$) is given by
\begin{equation}
i_{L} = I_{DS}(v_{DS2}, v_{GS2}, L-L_{eff}),
\label{Eq:eq18}
\end{equation}
\begin{equation}
v_{GS2} = V_{GS0},
\label{Eq:eq31}
\end{equation}
\begin{equation}
i_{L} = i_{DS2},
\label{Eq:eq32}
\end{equation}
\begin{equation}
i_{DS1} = i_{DS2}+i_{C},
\label{Eq:eq33}
\end{equation}
After averaging Eq.~\eqref{Eq:eq15} over the period $\tau$ and taking into account Eqs. \eqref{Eq:eq16}, \eqref{Eq:eq17}, \eqref{Eq:eq32}, \eqref{Eq:eq33} and $\langle i_C \rangle = 0$ we obtain equation for iL the current in the read out circuit:
$$
i_L=\frac{1}{\tau}\int_{0}^{\tau} I_{DS} [-v_{DS2}-Ri_L+$$
\begin{equation}
\label{Eq:eq20}
V_{tr}\cos{(\omega t)}, V_{GS0}+V_{tr}\cos{(\omega t)}, L_{eff}]dt,
\end{equation}

Measured voltage $\Delta U$ is
\begin{equation}
\Delta U=R_L i_L
\label{Eq:eq21}
\end{equation}

Eq.~\eqref{Eq:eq18} and~\eqref{Eq:eq21} forms system of nonlinear equation with respect to $v_{DS2}$ and $i_L$.  In the case of low radiation intensity ($V_{tr} \ll \varphi_{T}$) Eq.~\eqref{Eq:eq18},~\eqref{Eq:eq20},~\eqref{Eq:eq21} simplifies to formula Eq.~\eqref{Eq:eq8} but with a coefficient $\frac{1}{2}$ instead of $\frac{1}{4}$ and the photoresponse is proportional to the radiation power.  For large radiation intensity ($V_{tr} \geq \varphi_{T}$) numerical methods should be used.

For low input intensities the simple analytical model Eq.~\eqref{Eq:eq4} of the broadband photoresponse based on Dyakonov and Shur model~\cite{bib25} was proposed in Ref.~\cite{bib28}.  It allows calculating THz phoresponse using static I-V characteristics:
\begin{equation}
\Delta U =\frac{V_{tr}^2}{4} \eta_L F_{\sigma},
\label{Eq:eq8}
\end{equation}

where, $V_{tr}$ is the amplitude of the \textit{ac} voltage induced between the gate and source by the THz radiation, $F_{\sigma}$ is the function of the channel conductivity;
\begin{equation}
F_{\sigma} =[\frac{1}{\sigma_{ch}} \frac{d\sigma_{ch}}{dV_{GS}}]_{V_{DS} \rightarrow 0},
\label{Eq:eq9}
\end{equation}

$V_{GS}$ is the \textit{dc} voltage between the gate and source, $\sigma_{ch}$ is channel conductivity, $\eta_L$ is the voltage divider transfer coefficient:
\begin{equation}
\eta_L = \frac{1}{1+R_{ch}/Z_L} ,
\label{Eq:eq10}
\end{equation}
\begin{equation}
Z_L = R_L \| \frac{1}{j 2 \pi f_{mod} C_{load}},
\label{Eq:eq11}
\end{equation}

where, $R_{ch}$ is the channel resistance, $Z_L$ is the complex load impedance of the setup, $R_L$ and $C_{load}$ are the resistance and the capacitance of measurement circuit, $f_{mod}$ is the modulation frequency.  The Eq.~\eqref{Eq:eq10} takes into account the fact that the photoresponse curve $\Delta U$ depends on the load impedance value of the read-out circuit, since this load impedance forms the voltage-divider with the transistor channel .

The function $F_{\sigma}$ in Eq.~\eqref{Eq:eq8} is expressed through $\sigma_{ch}$, which is useful for processing experimental data.  When analytic current expression $I_{DS}(V_{DS}, V_{GS})$ for FET device is known, $F_{\sigma}$ (Eq,~\eqref{Eq:eq9}) can be written as:
\begin{equation}
F_{\sigma} =(\frac{\partial I_{DS}}{\partial V_{DS}})^{-1} \frac{\partial^2 I_{DS}}{\partial V_{GS} \partial V_{DS}},
\label{Eq:eq12}
\end{equation}

The Eq.~\eqref{Eq:eq8} relates photoresponse value with \textit{dc} characteristics of the transistor.  There is no frequency dependency in Eq.~\eqref{Eq:eq8} because it assumes broadband detection.  Generally, it should be corrected by introducing a power and frequency dependent factor that depends on detection mechanism of antenna and matching between antenna and channel~\cite{bib23, bib24} (see Eq.~\eqref{Eq:eq8}, Eq.~\eqref{Eq:eq6}).

Following Ref.~\cite{bib29} we write the drain-source current $I_{DS}(V_{DS}, V_{GS})$.
$$ I_{DS}(V_{DS}, V_{GS}, L)=\frac{W}{L} \mu_{n} C'_{ox} 2 \eta {\varphi_T}^2 \{[\text{ln}(1+$$
$$\text{exp}(\frac{V_{GS}-V_{th}+\alpha \eta V_{DS}}{2 \eta \varphi_T})]^2-$$
\begin{equation}
[\text{ln}(1+\text{exp}(\frac{V_{GS}-V_{th}-\beta \eta V_{DS}}{2 \eta \varphi_T})]^2\},
\label{Eq:eq13}
\end{equation}

where, \textit{W} and \textit{L} are width and length of transistor channel, respectively; $\mu_{n}$ is electron mobility in channel, $C'_{ox}$ is the gate oxide capacitance per unit area, $\varphi_{T}$ = $k_{B}T/q$ is the thermal voltage, $k_{B}$ is the Boltzmann constant, \textit{T} is the temperature, \textit{q} is the electron charge.
Eq.~\eqref{Eq:eq13} describes the channel current in all inversion and saturation regions~\cite{bib29}. It was used to fit experimental static $I-V$ data.  Fig.~\ref{Fig:fig04}(b) shows experimental and calculated $I-V$ characteristics (right ordinate) and photoresponses (left ordinate) for HEMT (HEMT parameters are in Table~\ref{tab:tab01}).  The Eq.~\eqref{Eq:eq13} was originally developed for silicon FET~\cite{bib29} but it describes HEMT data sufficiently good. With respect to the the original equation of Ref.~\cite{bib29} we added in Eq.~(\eqref{Eq:eq13}) terms containing coefficients \textit{$\alpha$} and \textit{$\beta $} (in original formula \textit{$\alpha$}~=~0, \textit{$\beta $}~=~1).  These parameters allows for better matching with experimental data and are related to presence of short-channel effects (slight current increase observed in saturation region).

It is worth to mention that most of  earlier developed models ~\cite{bib21, bib23, bib30} used expression of $I_{DS}(V_{DS},~V_{GS})$ valid either only in the weak inversion or only in the strong inversion regions.  Expression Eq.~\eqref{Eq:eq13} is phenomenological expression valid in whole (strong and weak inversion) ranges.

The high intensity radiation can result in carriers heating and in heating of transistor itself.  Temperature is one of the main parameters that influence on transistor characteristics.  Effective mobility decreases with temperature increase.  It changes characteristics of the transistor and thus changes its intrinsic responsivity and impedance matching between antenna and FET.  Earlier in Ref~\cite{bib34} was studied the photoresponse behavior for low temperatures, but the study of photoresponse behavior for temperatures higher than 300~K.  In strong-inversion regime the channel mobility in FET devices can be written as~\cite{bib29}.
\begin{equation}
\mu_{n}(T)=\mu_{n}(T_{r})(\frac{T}{T_{r}})^{-k_{\mu}},
\label{Eq:eq22}
\end{equation}

where $T_r$ is temperature at which parameters were extracted and $k_{\mu}$ is constant temperature coefficient, which varies between 1.2 and 2.~\cite{bib29, bib35}

The threshold voltage decreases linearly with temperature~\cite{bib29}
\begin{equation}
V_{th}(T)=V_{th}(T_{r})-k_{th}(T-T_{r}),
\label{Eq:eq23}
\end{equation}

where, $k_{th}$ for Si-MOSFET is between 0.5 and 3~mV/K~\cite{bib29}, for HEMT device $k_{th}$ is smaller for e.g. 0.3~mV/K.~\cite{bib36}  Results of simulation using Eq.~\eqref{Eq:eq22} and Eq.~\eqref{Eq:eq23} are presented in Fig.~\ref{Fig:fig04} (b).  One can see relatively good agreement of calculations with experimental data validating our model.

 To simulate the photoresponse following calculations steps are performed : i) transistor parameters are extracted from \textit{dc} measurements data (Fig.~\ref{Fig:fig04} and Fig.~\ref{Fig:fig06}(b)) (see Table~\ref{tab:tab01}), ii) Eq.~\eqref{Eq:eq20} is numerically solved, iii) $\Delta U$ is found using Eq.~\eqref{Eq:eq21}, iv) comparing experimental and model data in linear region, the constant $k_{ant}$ is determined and introduced into Eq.~\eqref{Eq:eq7}. The final simulation results of our model are presented in Fig.~\ref{Fig:fig06}(a).  The model takes into account the non-linear behavior of the current in the transistor channel.  One can see that our model fits well with the experimental data for MOSFET in all ranges; for HEMT there is a slight deviation from experiment at high intensities.  It is worth noting that the photoresponse model accuracy depends on the accuracy of the channel current model.  For MOSFETs the dc current is well described by Eq.~\eqref{Eq:eq13}.  However currently there is no a single well developed model of $I_{DS}(V_{DS},~V_{GS})$ which takes into account all effects corresponding to HEMT transistors.  Therefore, the accuracy of the experiment described by model Eq.~\eqref{Eq:eq13} for HEMT is slightly less than for MOSFET transistors.

Performing the simulations with difrent sets of parameters we have observed that in principle the saturation behaviuor can be also explained by transistor heating by incoming radiation. However we found that reproducing the saturation requires unrealistically high temperatures. Therefore we can with certainty state that the current saturation (similar to one observed in standard dc characteristics) is the dominating effect responsible for THz signal saturation at high radiation intensities.
%%%%%%%%%%%%
\section{Conclusions}
%%%%%%%%%%%%
To summarize, the detection by field effect transistors in broadband non-resonant regime was observed in wide intensity range from 0.5~mW/cm$^2$ up to 500~kW/cm$^2$ at the frequency range from 0.13~THz up to 3.3~THz.  We demonstrate that detection can be linear with respect to radiation intensity up to several kW/cm$^2$ and then it saturates.  The signal behavior in wide range of radiation intensity is interpreted in the frame of the generalized model of THz FET detection. This model takes into account the nonlinear behavior of the transistor characteristics which is important at high level of photoresponse.  The model quantitatively explain experimental data both in linear and nonlinear (saturation) range showing that dynamic range of field effect transistors based terahertz detectors extends over many orders of magnitude.
%%%%%%%%%%%%%%%%%%%%%%%%%%%%%%%%%%%%%%%%%%%%%%%%%%%%%%%%%%%%%%%%%%%%%%%%%%%%%%%%%%%%
\section{Acknowledgements}
This work was supported by ANR project ``WITH'' and by CNRS and GDR-I project ``Semiconductor sources and detectors of THz frequencies''. PUF project; DFG (SPP\~1459 and GRK\~1570), Linkage Grant of IB of BMBF at DLR. We are grateful to M.I. Dyakonov, K. Romanov, and M. Levinshtein for fruitful discussions.
%%%%%%%%%%%%%%%%%%%%%%%%%%%%%%%%%%%%%%%%%%%%%%%%%%%%%%%%%%%%%%%%%%%%%%%%%%%%%%%%%%%%

%%%%%%%%%%%%%%%%%%%%%%%%%%%%%%%%%%%%%%%%%%%%%%%%%%%%%%%%%%%%%%%%%%%%%%%%%%%%%%%%%%%%
\end{document}